\documentclass[10pt, a4paper]{article}

\usepackage[utf8]{inputenc}

\usepackage{amssymb}
\usepackage{amsmath}
\usepackage{amsthm}
\usepackage{adjustbox}

\usepackage[english]{babel}
\usepackage[backend=biber]{biblatex}

\usepackage{csquotes}

\usepackage[inline]{enumitem}

\usepackage{float}
\usepackage[T1]{fontenc}

\usepackage{graphicx}
\usepackage[automake]{glossaries-extra}
\setabbreviationstyle{long-em-short-em}

\usepackage{multicol}
\usepackage{multirow}

\usepackage{tikz}
\usetikzlibrary{arrows, shapes, automata, patterns, matrix, positioning}

\usepackage{wasysym}

\usepackage{bm}
\usepackage{accents}
\usepackage{authblk}

\input{lib/define}
\newabbreviation{mtd}{MTD}{Moving Target Defense}
\newabbreviation[plural={OSes}, firstplural={operating systems (OSes)}]{os}{OS}{operating system}
\newabbreviation{dag}{DAG}{Directed Acyclic Graph}
\newabbreviation{at}{AT}{Attack Tree}
\newabbreviation{adt}{ADT}{Attack Defense Tree}
\newabbreviation{cgs}{CGS}{Concurrent Game Structure}
\newabbreviation{amg}{AMG}{Attack Moving target defense DAG}
\newabbreviation{ptmdp}{PTMDP}{Priced Timed Markov Decision Process}
\newabbreviation{cdf}{CDF}{Cumulative Distribution Function}
\newabbreviation{pta}{PTA}{Priced Timed Automata}
\newabbreviation{ptg}{PTG}{Priced Timed Game}
\newabbreviation{mp}{MP}{moving parameter}
\newabbreviation{dlsef}{DLSeF}{Dynamic key-Length-based Security Framework}
\newabbreviation{dare}{DARE}{Dynamic Application Rotation Environment}
\newabbreviation{aslr}{ASLR}{Address Space Layout Randomization}
\newabbreviation{more}{MORE}{Multiple OS Rotational Environment}
\newabbreviation{iot}{IoT}{Internet of Things}
\newabbreviation{qos}{QoS}{Quality of Service}
\newabbreviation{nvd}{NVD}{National Vulnerability Database}
\newabbreviation{ads}{ADS}{Attack Defense tree with Sequential conjuntion}
\newabbreviation{sat}{SAND-AT}{Attack Tree with Sequential Conjuntion}
\newabbreviation{upgs}{CapCGS}{Capacity Concurrent Game Structure}
\newabbreviation{upatl}{CapATL}{Capacity Alternating-time Temporal Logic}
\newabbreviation{capsl}{CapSL}{Capacity Strategy Logic}
\newabbreviation{capcgs}{CapCGS}{Capacity Concurrent Game Structure}
\newabbreviation{capatl}{CapATL}{Capacity Alternating-time Temporal Logic}
\newabbreviation{atl}{ATL}{Alternating-time Temporal Logic}
\newabbreviation{atel}{ATEL}{Alternating-time Temporal Epistemic Logic}
\newabbreviation{sl}{SL}{Strategy Logic}
\newabbreviation{it}{IT}{Information Technology}
\newabbreviation{is}{IS}{Information Security}
\newabbreviation{cti}{CTI}{Cyber Threat Intelligence}
\newabbreviation{enisa}{ENISA}{European Union Agency for Cybersecurity}
\newabbreviation{mas}{MAS}{Multi-Agent System}
\newabbreviation{ctl}{CTL}{Computation Tree Logic}
\newabbreviation{ids}{IDS}{Intrusion Detection System}
\newabbreviation{sps}{SPS}{Self-Protecting Software}

\makeglossaries
\begin{document}

\title{\huge\bf Capacity ATL}
\author[1]{\footnotesize Gabriel Ballot}
\author[2]{Vadim Malvone}
\author[2]{Jean Leneutre}
\author[3]{Youssef Laarouchi}
\affil[1]{SEIDO Lab, EDF R\&D - Télécom Paris, Institut Polytechnique de Paris,
Palaiseau, France}
\affil[2]{LTCI, Télécom Paris, Institut Polytechnique de Paris, Palaiseau,
France}
\affil[3]{EDF R\&D Lab Saclay, Palaiseau, France}
\date{}

\maketitle

\begin{abstract}
  Model checking strategic abilities was successfully developed and applied since the
  early 2000s to ensure properties in \glsfmtlong{mas}. In this paper, we
  introduce the notion of \textit{capacities} giving different
  abilities to an agent. This applies naturally to
  systems where multiple entities can play the same role in the game, such as
  different client versions in protocol analysis, different robots in
  heterogeneous fleets, different
  personality traits in social structure modeling, or different attacker
  profiles in a cybersecurity setting. With the capacity of
  other agents being unknown at the beginning of the game, the longstanding
  problems of imperfect
  information arise. Our contribution is the following:
  \begin{enumerate*}[label=(\emph{\roman*})]
    \item we define a new class of concurrent game structures where the agents
      have different capacities that modify their action list and
    \item we introduce a logic extending \glsfmtlong{atl} to reason about these
      games.
  \end{enumerate*}
\end{abstract}

\section{Introduction}

Engineers design increasingly complex systems, and relying on expert knowledge
to trust that the system behaves as
expected would be illusional. To tackle this issue, researchers developed \textit{formal
verification techniques}
to rigorously prove some properties of systems' correctness. \textit{Model
checking}~\cite{Clarke2018Handbookmodelchecking} is the branch of formal verification that aims to check
specifications for all system computations. It relies on
three components: a modeling formalism to have a formal model of the real system, a
specification formalism to express non-ambiguous properties, and a
model-checking algorithm to verify if a given property holds on a given model.
A popular framework is \gls{ctl}~\cite{Emerson1982UsingBranchingTime} to
express properties on infinite computation trees
when the program is abstracted as a finite Kripke structure (transition
system).

Verification was applied to closed systems until the early 2000s. However, the
need to analyze open systems
appeared since now all systems are connected and distributed. As a result, 
the model-checking community gained interest in verifying
\glspl{mas} that model the interaction between different entities
that may, or may not, have the same objective. The Kripke structure is
replaced by \gls{cgs} where a tuple
of agent actions triggers transitions. Since then, many
theories~\cite{Peterson2002Decisionalgorithmsmultiplayer,
Alur2002AlternatingTimeTemporal, Mogavero2014ReasoningStrategiesModel} and
tools~\cite{David2015UppaalStratego, Lomuscio2009MCMASModelChecker,
Alur2001jMochamodelchecking, Ferrando2021StrategyRVTool, Niewiadomski2020MsATLToolSAT} to
reason about \glspl{mas} have
emerged, including
\gls{atl}~\cite{Alur2002AlternatingTimeTemporal}. \gls{atl} let us reason about
the strategic abilities of sets of agents called \textit{coalitions}. It can check
if a coalition has a strategy to enforce a property in a specific temporal
horizon, whatever the actions of agents outside the coalition. For example, the
\gls{atl} property $readCmd \rightarrow \strat{controler} (\neg write) \until
read$ could mean that ``if a \verb_read_ command arrives, the memory controller
can prevent any write in the register until the read happens''. \gls{atl}
extensions are regularly proposed to reason about more complex systems, for
instance, considering bounded resources~\cite{Alechina2010ResourceBoundedAlternating,
Nguyen2019ProbabilisticResourceBounded},
probability~\cite{Chen2007ProbabilisticAlternatingTime, Nguyen2019ProbabilisticResourceBounded}, discrete
time~\cite{Andre2017TimedATLForget}, different strategic
abilities~\cite{Aagotnes2007AlternatingTimeTemporal,
Mogavero2014ReasoningStrategiesModel, Walther2007AlternatingTimeTemporal}, or imperfect
information~\cite{Hoek2002TractableMultiagentPlanning,
Jamroga2004AgentsThatKnow, Dima2010ModelCheckingAlternating}.

\paragraph{Contribution}
This paper extends \gls{atl} with the new notion of 
agents' \textit{capacities}. At the beginning of the game, each agent is
assigned (or chooses) secretly a capacity they will keep during the
game. This capacity
determines the set of actions available for the agent. The agent capacity diversity
is very intuitive since, in many real-life situations, different
entities can take the role of an agent. For instance,
considering sports, the opponent may be right-handed or left-handed, and the
actions of a right-handed player may not be the same as the left-handed player.
When the match starts, the agents do not know if the opponent is right or
left-handed, but they might identify it during the game and act consequently.
More practically, we see many scenarios where agents can have different
capacities influencing their actions and where their capacities are not
publicly known. In distributed computing, an agent may run one among
different protocol versions and not declare it publicly. Finding a distributed protocol
verifying good properties where protocol versions are uncertain is not easily modeled for
now. In contrast, it could be modeled using one capacity per client version in
our setting.
A fleet of heterogeneous robots could be modeled in our framework with a
capacity per type of robot. In social structure modeling, the different 
personality traits of agents can be capacities: some people are altruists,
adventurous, selfish, \etc.
In cyber security, the attacker may also have different
capacities
corresponding to his resources and skills. 
Identifying the attacker's capacity and responding accordingly is a fundamental and
challenging problem in cyber security.
The contributions of this paper are the following:
\begin{enumerate*}[label=(\emph{\roman*})]
  \item we define a new class of concurrent game structures, called \gls{upgs}, where the agents
    have different capacities that modify their action list, and
  \item we introduce \gls{upatl}, a logic extending \glsfmtlong{atl} to reason
    about \glspl{upgs}.
\end{enumerate*}

\paragraph{Related work}
In the early 2000s, Hoek \etal introduced an epistemic extension of
\gls{atl}~\cite{Hoek2002TractableMultiagentPlanning}, leading to a series of
works on imperfect-information strategic logics like \gls{atel}~\cite{Jamroga2004AgentsThatKnow,
Dima2010ModelCheckingAlternating, Dima2011ModelCheckingATL}. In
these settings, the agents cannot
distinguish some system states, making it harder to find a winning
strategy (three-player perfect recall reachability is enough to
make the problem undecidable~\cite{Dima2011ModelCheckingATL}).
In~\cite{Malvone2017HidingActionsMulti}, agent actions' indistinguishability
translates to \textit{imperfect information} on states.
The recent focus was highlighting \gls{atl} with imperfect information
fragments to retrieve decidability~\cite{Berthon2017DecidabilityResultsATL*} or
find approximative verification
algorithms~\cite{Jamroga2019ApproximateVerificationStrategic,
Belardinelli2023AbstractionRefinementFramework, Belardinelli2022ApproximatingPerfectRecall}. Our logic
introduces imperfect information on agent capacities, which is not
considered in current imperfect information games. Another type of game, called
\textit{incomplete information}
games~\cite{Harsanyi1967GamesIncompleteInformation}, makes it possible to have different player
profiles and agents are unaware of opponents' profiles. The information level
of this type of game is more related to \gls{upatl}'s imperfect information on
agent capacities. Incomplete information games have been applied in the cyber security
context~\cite{Roy2010SurveyGameTheory}. In~\cite{Liu2006BayesianGameApproach},
optimal \gls{ids} allocation is derived from a Bayesian game where a node can
have a \textit{malicious} or \textit{normal} profile, leading to different
payoff functions. In~\cite{EmamiTaba2016BayesianGameDecision}, the defender
tries to keep security goals secret in \glsfmtlong{sps} systems and discover the adversary type and
objective thanks to a Bayesian game. However, incomplete information games
are based on reward functions and equilibria rather than a game structure and a
logic. Thus, the analysis stands for one- and not multi-step attacks, as in our
setting.

\section{\glsfmtlong{upgs}}\label{sec:game-structure}

In this section, we define the \glsfmtfull{upgs}, an extension of \gls{cgs}, to
consider the possibility that agents have different \textit{capacities}
restricting their actions. First, we highlight some general notations.

\paragraph{Notations}
Given a set $X$, $\parts{X}$ denotes the set of subsets of $X$.
We write $f: X \pmapsto Y$ (resp. $g: X \to Y$) to introduce a partial function
$f$ (resp. application $g$) from a set $X$ to a set $Y$, and
$\dom(f)$ denotes the domain of $f$. For an integer $n>0$, we denote the set
$\{1, \dots, n\}$ by $\integerSet{i}$, and $\Nn$ is the set of all the
positive integer. For a subset $I = \{i_1, i_2, \dots\} \subseteq \Nn$,
we denote $2I = \{2i_1, 2i_2, \dots\}$ and $I-1 = \{i_1 -1, i_2-1, \dots\}$.
Given a sequence of elements $s = s_1s_2\dots$, we denote the length of $s$
(possibly infinity) by $\vert s\vert$, and we denote the subsequence $s_I =
(s_i)_{i\in I}$. As such, the $i^{th}$ element of $s$ is
$\access{s}{i} = s_i$ and the prefix of length $i$ of $s$ is $\prefix{s}{i}$.
Moreover, the last element of $s$ (if $\vert s\vert< \infty$) is denoted by
$\last{s}$.
Finally, we use $\top$ to denote the value \textit{true} and $\bot$ to denote the value
\textit{false}.

\begin{definition}[\glsfmtlong{upgs}] 
  A \gls{upgs} is a
  structure $\upgs = \upgsDef$ with the following attributes: 
  a set of $\nbPlayers$ \textit{agents} $\Players = \{1, \dots, \nbPlayers\}$,
  a finite set of \textit{capacities} $\Capacities$, 
  a finite set of \textit{states} $\states$, 
  a finite set of \textit{atomic propositions} $\labels$, 
  a \textit{labeling function} $\labeling: \states \to \parts{\labels}$, 
  a finite set of \textit{actions} $\Actions$, 
  a function $\PlayerCapacitiesMapDef: \Players \to
  \parts{\Capacities}$ that assigns a subset of capacities to each agent,
  a function $\capacityActionMapDef: \Capacities \to
  \parts{\Actions}$ that assigns a subset of actions to each capacity, 
  a list $\protocolDef = (\protocolDef_\player)_{\player\in \Players}$ of
  \textit{protocol functions} where $\protocolDef_{\player}: \states \to
  \parts{\Actions}$ gives the set of actions $\protocol{\player}{\state}$,
  available for the agent $\player \in\Players$ in
  the state $\state$ and verifies $\protocol{\player}{\state} \subseteq
  \bigcup_{\capacity \in
  \PlayerCapacitiesMap{\player}}\capacityActionMap{\capacity}$ and
  $\protocol{\player}{\state} \cap
  \capacityActionMap{\capacity} \neq \emptyset$ for all $\capacity \in
  \PlayerCapacitiesMap{\player}$,
  and a partial \textit{transition function} $\transition: \states \times
  \Actions^\nbPlayers \pmapsto 
  \states$ defined for all $(\state, \action_1, \dots, \action_\nbPlayers)$ verifying
  $\action_\player \in \protocol{\player}{\state}$ for all $\player \in \Players$.
\end{definition}

\begin{remark}\label{rk:protocol}
  The restriction on the protocol function, called \textit{progression
  condition}, imposes that every agent, whatever
  its capacity, has at least one action
  available, and a transition exists whatever the combination of available
  actions of individual agents.
\end{remark}

We consider a general $\gls{upgs}$ $\upgs =
\upgsDef$ with $\nbPlayers$ agents in the rest of this paper.
Intuitively, during a play, each agent $\player \in \Players$ will secretly
choose one of its 
capacities $\capacity \in \PlayerCapacitiesMap{\player}$, meaning that he can
use only the actions $\action \in \capacityActionMap{\capacity}$.
An interesting question to ask (and we will formalize it in
Section~\ref{sec:logic}) is whether a coalition has a strategy to guarantee
properties, including identifying the capacity of other agents.
First, we remind some definitions that apply to \gls{cgs} and
\gls{upgs}. A \textit{path} $\computation$ describes the possible realizations of the
game. It contains information about the succession of states and actions of
all the agents. It is formalized as a possibly
infinite sequence $\state_1\actionVector_1\state_2\actionVector_2\dots$ often
written
$\computation = \state_1 \xrightarrow{\actionVector_1} \state_2
\xrightarrow{\actionVector_2} \dots$
where $\actionVector_i = (\action_i^1, \dots, \action_i^\nbPlayers) \in
\Actions^\nbPlayers$ is the agents joint
action at step $i$. It must satisfy for all $i$,
$\state_{i+1} = \transition(\state_{i}, \action_i^1, \dots, \action_i^\nbPlayers)$. 
If a path is finite, it ends with a state.
The set of paths is denoted by $\Paths$.
Given a path $\computation = \state_1 \xrightarrow{\actionVector_1} \state_2
\xrightarrow{\actionVector_2} \dots$, 
the \textit{state trace} is the sequence of states $\computation_{2\Nn - 1} =
\state_1\state_2\dots$ and the \textit{action trace} is the sequence of
joint actions $\computation_{2\Nn} =
\actionVector_1\actionVector_2\dots$. Finally,
we call \textit{history} a finite state trace and we let $\Histories =
\{\StateTrace{\computation}
\mid \computation \in \Paths, \vert \computation\vert < \infty\}$ be the set of
histories.

As \gls{atl}, \gls{upatl} is a logic to reason about \textit{strategies}, so we 
formalize this concept as a mapping between histories and actions (such
strategies are also called \textit{memoryful strategies}).
Moreover, we will use \textit{assignment} functions similarly to~\cite{Mogavero2014ReasoningStrategiesModel}
to assign a strategy or a capacity to an agent.
Notice that if a memoryful strategy satisfies an \gls{atl} property, then a
memoryless strategy (mapping only states to the actions)
also exists to satisfy this property. Consequently, \gls{atl} reasoning needs
only to deal with memoryless strategies. However, this is not true in
\gls{upatl}, and we have to reason about the memoryful class of
strategy.

\begin{definition}[Strategy Assignment]
  A \textit{strategy} is a function $\strategy: \Histories
  \to \Actions$ that maps each history to an action.
  A \textit{strategy assignment} is a partial function
  $\assignStrategy:
  \Players \pmapsto (\Histories \to \Actions)$ that assigns strategies to
  agents. It must verify, for $\player \in \dom(\assignStrategy)$ and $\history
  \in \Histories$, that $\assignStrategy(\player)(\history)
  \in \protocol{\player}{\last{\history}}$.
  We denote by $\assignStrategySet$ the set of strategy assignments and by
  $\assignStrategySetFor{\playerCoalition}$ the set of strategy assignment with
  domain $\playerCoalition \subseteq \Players$.
\end{definition}

\begin{definition}[Capacity Assignment]
  A \textit{capacity assignment} is a partial function $\assignCapacity:
  \Players \pmapsto \Capacities$ that assigns capacities to agents, \st, for an
  agent $\player \in \Players$, we have $\assignCapacity(\player) \in
  \PlayerCapacitiesMap{\player}$.
  We use $\assignCapacitySet$ to denote the
  set of capacity assignments and
  $\assignCapacitySetFor{\playerCoalition}$ to denote the set of
  capacity assignments with
  domain $\playerCoalition \subseteq \Players$.
\end{definition}

We say that a (capacity or strategy)
assignment $\assign$ is complete if it is defined for all agents,
\ie, $\dom(\assign) = \Players$.
In the \gls{upgs}, the agents are assigned a capacity once and for all.
Consequently,
given a path $\computation$, we can rule out the complete capacity assignments that do
not authorize the agent to use some actions from $\ActionTrace{\computation}$. We formalize this in the
notion of $\computation$-\textit{compatible assignments}.
Given a path $\computation$, we let
$\capacityTransitionFunction(\computation) \subseteq
\assignCapacitySetFor{\Players}$ denote the set of possible complete capacity
assignments that may bring about $\computation$.
We have $\capacityTransitionFunction(\computation) = \{\assignCapacity \in
\assignCapacitySetFor{\Players} \mid \forall \actionVector \in
\ActionTrace{\computation}, \forall \player \in \Players,
\access{\actionVector}{\player} \in
\capacityActionMap{\assignCapacity(\player)}\}$.
We can now define the outcomes of a strategy assignment
from a given finite path. It returns the extending paths respecting a capacity
assignment and the input strategy assignment.

\begin{definition}[Outcomes]\label{def:outcomes}
  Let $\computation = \state_1 \xrightarrow{\actionVector_1} \state_2
  \xrightarrow{\actionVector_2} \dots \state_j$ be a finite path and $\assignStrategy \in
  \assignStrategySetFor{\playerCoalition}$ be a partial strategy
  assignment for some coalition
  $\playerCoalition \subseteq \Players$.
  The set of \textit{outcomes}
  $\Outcomes(\computation, \assignStrategy) \subseteq \Paths$ is a set of infinite paths
  $\computation' = \computation \xrightarrow{\actionVector_{j}} \state_{j+1}
  \xrightarrow{\actionVector_{j+1}} \dots$ verifying
  $\capacityTransitionFunction(\computation') \neq \emptyset$, and,
  for all $i \geq j$ and $\player \in \playerCoalition$,
  $\access{\actionVector_i}{\player} =
  \assignStrategy(\player)(\state_j\dots\state_i)$.
\end{definition}

\section{\glsfmtlong{upatl}}\label{sec:logic}

This section introduces \gls{upatl}, an extension of \gls{atl}.
Subsection~\ref{sec:syntax} defines \gls{upatl} syntax, and
Subsection~\ref{sec:semantics} gives its semantics.

\subsection{Syntax}\label{sec:syntax}

We define a new logic extending \gls{atl}~\cite{Alur2002AlternatingTimeTemporal} called
\gls{upatl} to reason about unknown capacities. It can specify properties like
``Can a coalition of agents guess other agents' capacities?''.

\begin{definition}[\glsfmtlong{upatl} syntax]\label{def:syntax}
  The following grammar defines a \gls{upatl} formula $\phi$:
  \begin{align*}
    \phi &:== \ell \mid \knows{\player}{\varphi} \mid \neg \phi \mid \phi\land\phi
    \mid\strat{\playerCoalition}\psi \\
    \psi &:== \nnext \phi \mid \phi \until \phi \mid \phi \releases \phi \\
    \varphi &:== \hasCapacity{\player}{\capacity} \mid \neg \varphi \mid \varphi
    \land \varphi
  \end{align*}
  where $\ell \in \labels$ is an atomic proposition,
  $\playerCoalition \subseteq \Players$ is an agent coalition, $\player \in
  \Players$ is an agent, and $\capacity \in \Capacities$ is a capacity.
\end{definition}

As in \gls{atl}, $\strat{\cdot}$ is the \textit{strategic operator}, and
$\strat{\playerCoalition}\psi$ means that $\playerCoalition$ has a strategy to
enforce $\psi$ whatever the actions of the other agents. The strategic operator
is immediately followed by a \textit{temporal operator}, either $\nnext$ for
``next'', $\until$ for ``until'', or $\releases$ for ``release'' (the dual
of $\until$). Finally, $\knowsDef^{\player}$ is the \textit{knowledge
operator} and $\knows{\player}{\varphi}$ means that agent $\player$ knows
that the capacity assignment verifies $\varphi$ where $\varphi$ is called
\textit{capacity assignment formula}: it
characterizes a set of capacity assignments. For
example $\hasCapacity{\player_1}{\capacity_1} \land
(\hasCapacity{\player_2}{\capacity_2} \lor
\hasCapacity{\player_2}{\capacity_3})$ characterizes the set of complete
capacity assigments $\assignCapacity$ verifying $\assignCapacity(\player_1) =
\capacity_1$ and $\assignCapacity(\player_2) \in \{\capacity_2, \capacity_3\}$.
We also call \textit{path formula} a formula that derives form $\phi$ in
\gls{upatl} syntax. Note that \gls{upatl} formulae are the path formulae.

\begin{remark}
  It is noteworthy that our knowledge operator $\knowsDef^{\player}$ 
  deals with $\player$'s knowledge about agents' capacity while the usual knowledge operator
  in epistemic logics deals with properties of the model (\eg, ``Do agents know that
  property $\phi$ holds?'').
\end{remark}

\subsection{Semantics}\label{sec:semantics}

We consider that an agent $\player$ can only observe the succession of
states and the actions he did. This justifies the introduction of the following 
indistinguishability relation over $\Paths$ for each agent.

\begin{definition}[Indistinguishability]\label{def:indistinguishability}
  Let $\computation, \computation' \in \Paths$ be two paths.
  We say that $\computation$ and $\computation'$
  are indistinguishable for agent $\player \in \Players$,
  denoted by $\computation \sim_\player \computation$, iff 
  \begin{enumerate*}[label=(\emph{\roman*})]
      \item $\StateTrace{\computation} = \StateTrace{\computation'}$ and
      \item agent $\player$'s actions are the
        same, \ie, for any index $i \geq 0$, 
        $\access{\actionVector}{\player} =
        \access{\actionVector'}{\player}$, where
        $\actionVector = \access{\computation}{2i}$ and $\actionVector' =
        \access{\computation'}{2i}$.
  \end{enumerate*}
\end{definition}

\gls{upatl} semantics is formalized through a satisfaction relation for an
infinite path, an index of this path, and a capacity assignment.
Note that \gls{atl} uses state traces instead of paths because the actions between
states do not matter. However, in \gls{upatl}, having the actions is essential 
to determine what agents know.

\begin{definition}[\gls{upatl} semantics]\label{def:semantics}
  Let $\computation$ be an infinite path, $i > 0$ be an index, $\assignCapacity$
  be a complete capacity assignment, $\ell$ be an atomic proposition,
  $\player$ be an agent, $\playerCoalition$ be a coalition of agents, $(\phi,
  \phi_1, \phi_2)$ be three path formulae, $\psi$ be a temporal
  formula, and $(\vartheta, \vartheta_1, \vartheta_2)$ be three path or capacity
  assignment formulae.
  \gls{upatl} semantics is defined through the following satisfaction relation:
  \begin{itemize}
    \item $(\computation, i, \assignCapacity) \models \ell$ iff $\ell \in
      \labeling(\access{\computation}{2i-1})$,
    \item $(\computation, i, \assignCapacity) \models \hasCapacity{\player}{\capacity}$ iff
      $\assignCapacity(\player) = \capacity$,
    \item $(\computation, i, \assignCapacity) \models \knows{\player}{\varphi}$
      iff,
      for all $\computation' \sim_\player \computation$
      and $\assignCapacity' \in
      \capacityTransitionFunction(\prefix{\computation'}{2i-1})$,
      we have $(\computation', i, \assignCapacity') \models \varphi$,
    \item $(\computation, i, \assignCapacity) \models \neg \vartheta$ iff
      $(\computation, i, \assignCapacity) \not\models \vartheta$,
    \item $(\computation, i, \assignCapacity) \models \vartheta_1 \land
      \vartheta_2$ iff
      $(\computation, i, \assignCapacity) \models \vartheta_1$ and
      $(\computation, i, \assignCapacity) \models \vartheta_2$,
    \item $(\computation, i, \assignCapacity) \models \strat{\playerCoalition}\psi$, iff
      there is a strategy assignment $\assignStrategy$ for
      $\playerCoalition$\textemdash called winning strategy\textemdash such that $\Outcomes(\prefix{\computation}{2i-1}, \assignStrategy)
      \neq \emptyset$ and for any outcome $\computation' \in
      \Outcomes(\prefix{\computation}{2i-1}, \assignStrategy)$, we have
      $(\computation', i, \assignCapacity) \models \psi$,
      \ie,
      \begin{equation*}
        \exists \assignStrategy \in
        \assignStrategySetFor{\playerCoalition},
        \Outcomes(\prefix{\computation}{2i-1}, \assignStrategy) \neq \emptyset
        \land
        \forall \computation' \in \Outcomes(\prefix{\computation}{2i-1},
        \assignStrategy),
        (\computation', i, \assignCapacity) \models \psi
      \end{equation*}
    \item $(\computation, i, \assignCapacity) \models \nnext \phi$ iff
      $(\computation, i+1, \assignCapacity) \models \phi$,
    \item $(\computation, i, \assignCapacity) \models \phi_1 \until
      \phi_2$ iff there exists $j_1 \geq i$, \st,
      we have $(\computation, j_1, \assignCapacity) \models \phi_2$ and
      for all $i \leq j_2 < j_1$,
      we have $(\computation, j_2, \assignCapacity) \models \phi_1$,
    \item $(\computation, i, \assignCapacity) \models \phi_1 \releases
      \phi_2$ iff either
      \begin{enumerate*}[label=(\emph{\roman*})]
        \item for all $j \geq i$, we have $(\computation, j, \assignCapacity)
          \models \phi_2$, or
        \item there exisits $j_1 \geq i$, \st, 
          $(\computation, j_1, \assignCapacity) \models \phi_1 \land \phi_2$ and
          for all $i \leq j_2 < j_1$,
          we have $(\computation, j_2, \assignCapacity) \models \phi_2$.
      \end{enumerate*}
  \end{itemize}
\end{definition}

We say that a state $\state$ verifies a property $\phi$, denoted $\state
\models \phi$ iff there is a path $\computation$ and a complete capacity assignment
$\assignCapacity$ such that
$\access{\computation}{1} = \state$ and $(\computation, 1, \assignCapacity)
\models \phi$. Notice that, if such a $\computation$ and $\assignCapacity$ exist,
then all computation such that $\access{\computation}{1} = \state$ and
$\assignCapacity\in \assignCapacitySetFor{\Players}$ verify
$(\computation, 1, \assignCapacity) \models \phi$.

\begin{remark}
  The semantics of $\strat{\playerCoalition} \psi$ has an existential
  quantification over the strategy assignments for $\playerCoalition$ and a
  universal quantification over the outcomes. It implies
  an implicit existential (resp. universal) quantification over
  $\playerCoalition$ (resp. $\Players\setminus \playerCoalition$) capacity
  assignments compatible with past and future actions in each
  outcome.
  The compatibility with past actions comes from the fact that
  $\Outcomes(\computation, \assignStrategy)$ contains computations extending
  $\computation$. 
  Instead, we could have decided to extend only the last state
  $\last{\computation}$ which would authorize agents to pick a new
  capacity. Both choices are interesting, so we tackled the one that seems
  harder for model checking, and let the other option as a future extension.
\end{remark}

\section{Conclusion}\label{sec:conclusion}

This paper presented a concurrent game structure extension to include
different profiles for the agents to account for the diversity of entities
that may play the role of an agent. We described a logic extending \gls{atl} to
reason about the multiple agent capacity profiles.

\section*{Acknowledgement}

This work was carried out within SEIDO Lab, a joint
research laboratory covering research topics in the field of smart grids,
\eg, distributed intelligence, service collaboration, cybersecurity, and
privacy. It involves researchers from academia (Télécom Paris, Télécom
SudParis, CNRS LAAS) and industry (EDF R\&D).

{\footnotesize
\printbibliography
}

\end{document}